\documentclass[journal]{IEEEtran}

\usepackage[caption=false]{subfig}
\usepackage{mwe}
\usepackage{paralist}
\usepackage{balance}
\usepackage[printonlyused]{acronym}

\begin{document}


\title{Understanding the Modeling of Computer Network Delays using Neural Networks}

\author{Albert Mestres, Eduard Alarc\'on, Yusheng Ji and Albert Cabellos-Aparicio
\thanks{A. Mestres,  and A. Cabellos-Aparicio are with the Computer Architecture Department, Universitat Polit\`ecnica de Catalunya (e-mail: \{amestres/acabello\}@ac.upc.edu)}
\thanks{E. Alarc\'on is with the  Electrical Engineering Department, Universitat Polit\`ecnica de Catalunya (e-mail: eduard.alarcon@upc.edu)}
\thanks{Yusheng Ji is with NII, Japan (e-mail: kei@nii.ac.jp)}}

\maketitle

\begin{abstract}
Recent trends in networking are proposing the use of Machine Learning (ML) techniques for the control and operation of the network. In this context, ML can be used as a computer network modeling technique to build models that estimate the network performance. Indeed, network modeling is a central technique to many networking functions, for instance in the field of optimization, in which the model is used to search a configuration that satisfies the target policy.  
In this paper, we aim to provide an answer to the following question: \emph{Can neural networks accurately model the delay of a computer network as a function of the input traffic?} For this, we assume the network as a black-box that has as input a traffic matrix and as output delays. Then we train different neural networks models and evaluate its accuracy under different fundamental network characteristics: topology, size, traffic intensity and routing. With this, we aim to have a better understanding of computer network modeling with neural nets and ultimately provide practical guidelines on how such models need to be trained.

\end{abstract}



\begin{IEEEkeywords}
KDN, SDN, ML, Networking, Modeling
\end{IEEEkeywords}


\section{Introduction}

The use of Machine Learning (ML) techniques in the networking field is gaining momentum. One of the more promising areas is to help and improve the control and operation of computer networks. Although this idea is not new (see \emph{D.Clark et al.}) \cite{clark} this trend is becoming more popular thanks to two enabling technologies: Software-Defined Networking (SDN)~\cite{sdn} and Network Analytics (NA)~\cite{analytics}. 

Indeed, the rise of SDN transforms the network from an inherently distributed system to a (logically) centralized one that can be fully controlled through the SDN controller. At the same time, the NA field is developing techniques to monitor and obtain precise metrics of the network behavior. When combined, SDN and NA provide a central entity that offers a rich view and full control over the network where to apply ML.

In this context, learning techniques can be used to provide automatic control of the network via the SDN controller thanks to the network monitoring information obtained via the NA platform. This new networking paradigm is known as Knowledge-Defined Networking (KDN) \cite{KDN}.

Under the KDN paradigm, there are a wide variety of use-cases for taking advantage of ML techniques in computer networks. Among all such potential use-cases in this paper, we focus on a single one: \emph{modeling of network delays using neural networks(NNs)}. The main reason for this is that network modeling is central to many network operations, particularly in the field of network optimization. Typically network optimization algorithms require a network model over which the optimization techniques operate to find the best element (e.g.,~\cite{netOptimization2,netOptimization3}).

In this paper we aim to answer the following question: \emph{Can neural networks (NNs) accurately model the delay of a computer network as a function of the input traffic?} For this, we assume the network as a black-box that has as input traffic and as output delays. 
This question is fundamental to network modeling. Indeed both analytical (e.g., queuing models) and computational models (e.g., simulators) are well-known techniques used to estimate the performance of a network based on its input traffic. In this paper we posit that NNs can represent a third pillar in the area of network modeling, providing relevant advantages towards traditional techniques. Indeed with NNs, we can build a `digital twin' of the real infrastructure, this twin can then be used for optimization, validation, prediction, etc. To the best of our knowledge, this is the first attempt to model a computer network using NNs. 

Following this approach, in this paper, we first design a set of synthetic experiments and use different hyper-parameters and computer networks to understand how accurate are NNs when modeling computer networks. With this, we learn a set of practical guidelines that help us understand how NNs model computer networks. Finally, we validate our guidelines by effective modeling with a NN a realistic network loaded with realistic traffic matrices.

This article is organized as follows. Section~\ref{sec:useCase} presents a use-case in which modeling the delay of a computer network allows to optimize its performance. In Section~\ref{sec:pStatement}, we model the specific problem we address in this paper.  The related state of the art is presented in Section~\ref{sec:soa}. In Section~\ref{sec:methodology}, we introduce the methodology we followed to obtain the different datasets from simulated networks and to train different NN models. The different experimental results for different networks and NN models are presented in Section~\ref{sec:experiments}. In Section~\ref{sec:realistic}, we apply the methodology described in this paper to model a larger and more realistic network. Finally, in Section~\ref{sec:discussion} and Section~\ref{sec:conclusions}, we analyze the different experiments results obtained and conclude this work.

\section{A Use-Case of ML for Network Control and Operation}
\label{sec:useCase}

Network modeling is a well-established field that provides techniques which are central to a wide range of communication functions. Typically, techniques such as simulation and analytical tools are used to build computer network models. 

In our use-case, the network model is built \emph{using neural networks (NNs)}, specifically trained from collected or network simulated data. This model can be understood as a `digital twin' of the real networking infrastructure. As such it captures the fundamental relationship between network parameters, for instance, it can model the function that relates traffic load, routing policy with the resulting performance. 

\begin{figure}
  \centering
  \includegraphics[width=0.75\linewidth]{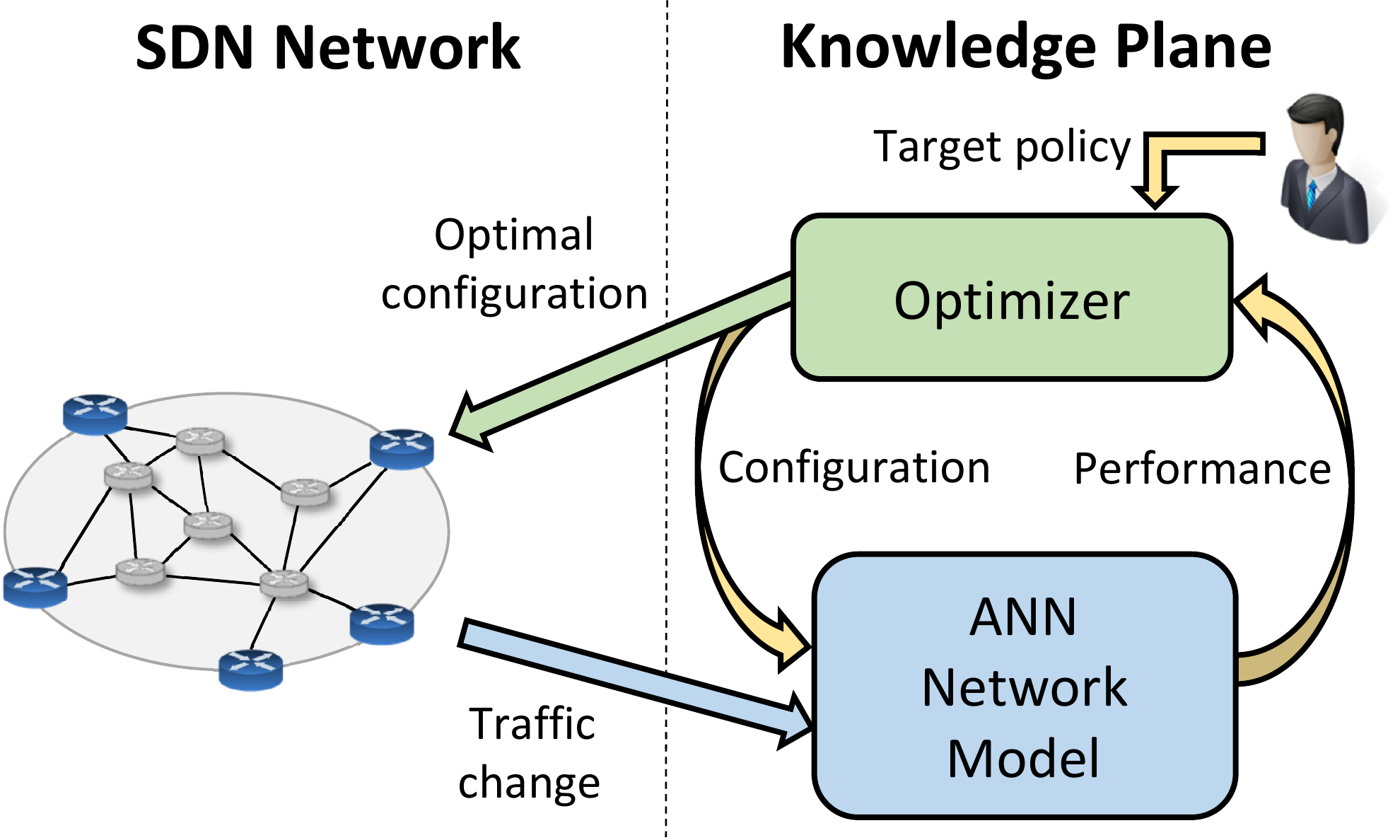}
  \caption{Graphical representation of the optimization use-case.} 
  \label{fig:usecase}
  \vspace{-0.25cm}
\end{figure}

As an example and in the context of network optimization, the `digital twin' can be used to estimate the performance of any possible incoming traffic and configuration (see ~Fig.~\ref{fig:usecase} for a schematic representation). Then traditional optimization algorithms (such as hill-climbing) can be used in combination with the `digital twin' to find the optimal configuration that results in the desired performance. Beyond optimization, the `digital twin' can also be used for validation, prediction, recommendation, etc.

In the context of this general use-case, this paper aims to have a better understanding on how NNs can learn from network data, how accurate they are, which are the fundamental challenges and ultimately to provide practical guidelines. 

\section{Problem Statement}
\label{sec:pStatement}

\begin{figure}
  \centering
  \includegraphics[width=\linewidth]{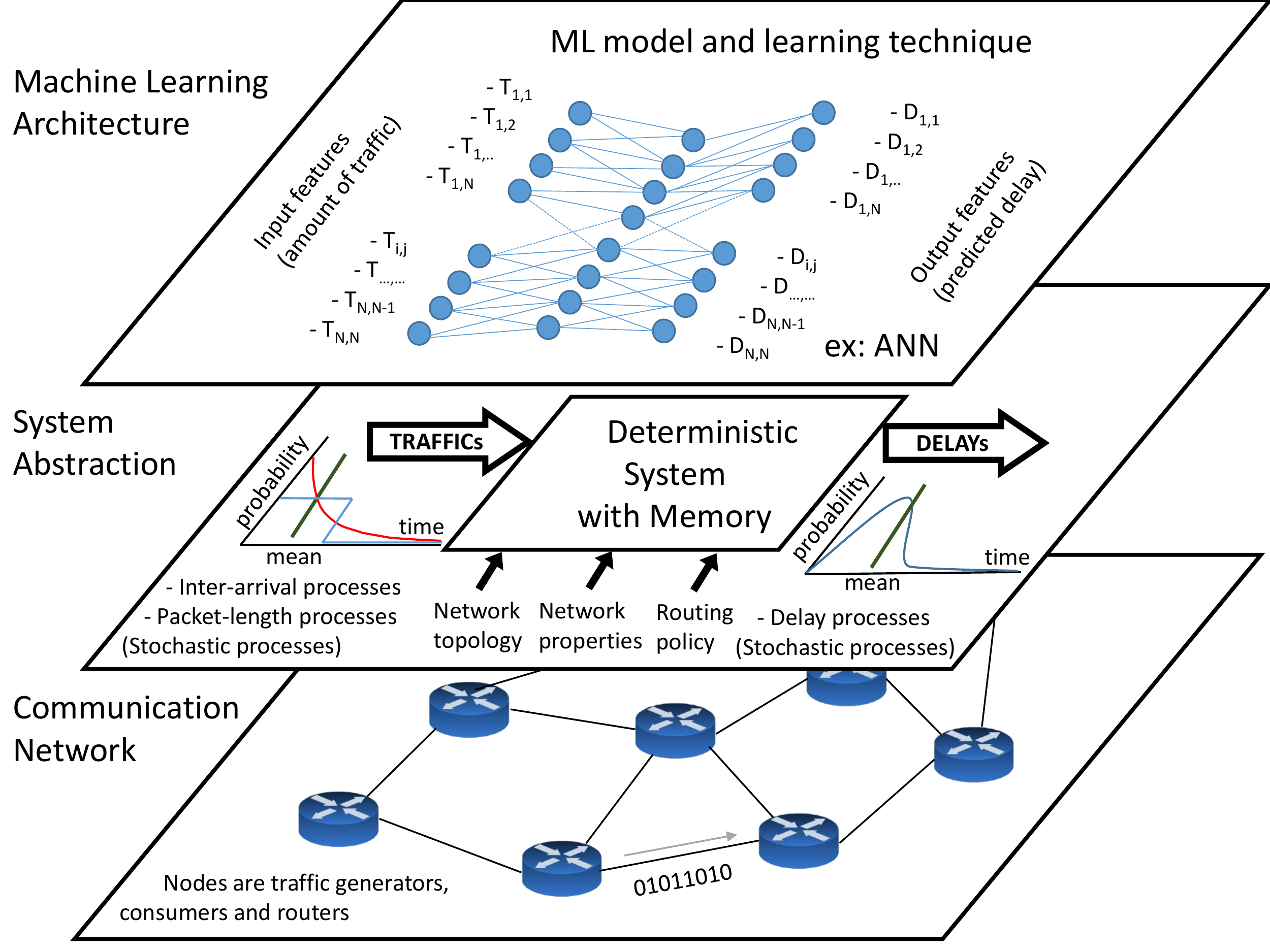}
  \caption{Graphical representation of the problem statement addressed in this paper.} 
  \label{fig:view}
  \vspace{-0.20cm}
\end{figure}

In this section, we describe the problem statement that we aim to address in this paper. Figure~\ref{fig:view} summarizes the problem statement using three layers. Note that in this paper, as a first step to understand how can NNs model the behavior of the network, we consider the configuration as constant and we build the delay model only as a function of the traffic.

The bottom layer represents the real-world physical network infrastructure that has certain fundamental characteristics, such as topology, size, routing, etc.
The middle layer represents the system abstraction in which the network is assumed as a black-box, traffic ingresses the box and egresses it with a certain average delay. The traffic is described by stochastic distributions, both the inter-arrival process and the packet length process. These stochastic processes are combined in the network, which is a deterministic complex system with certain properties (topology, routing, etc) and memory when random processes such as physical errors are not taken into account.

Finally, the top layer represents the NN that models the computer network performance. The NN produces estimates of the average end-to-end delay for all pairs of nodes considering the input traffic as a traffic matrix [ingress, egress]. The network characteristics (routing, topology, etc) are hidden from the NN, and hence it is trained only for one particular configuration of the network infrastructure that is, a certain topology, routing, etc. 

Specifically, the function we aim to model can be expressed as:
\vspace{-0.10cm}
$$
\mathbf{D} = f(\mathbf{T})
$$
\vspace{-0.10cm}
in which $\mathbf{D}$ and $\mathbf{T}$ are $N\times N$ matrices, the first one representing the average delay between the $i$ (row) node and the $j$ (column) node, and the second one representing the amount of traffic between $i$ and $j$ in an $N$ nodes network.
The delay from $i$ to $j$ is determined by the quantity of traffic sent between these nodes and the quantity of traffic sent between other nodes that share part of the path. 

\begin{figure}
  \centering
  \includegraphics[width=0.70\linewidth]{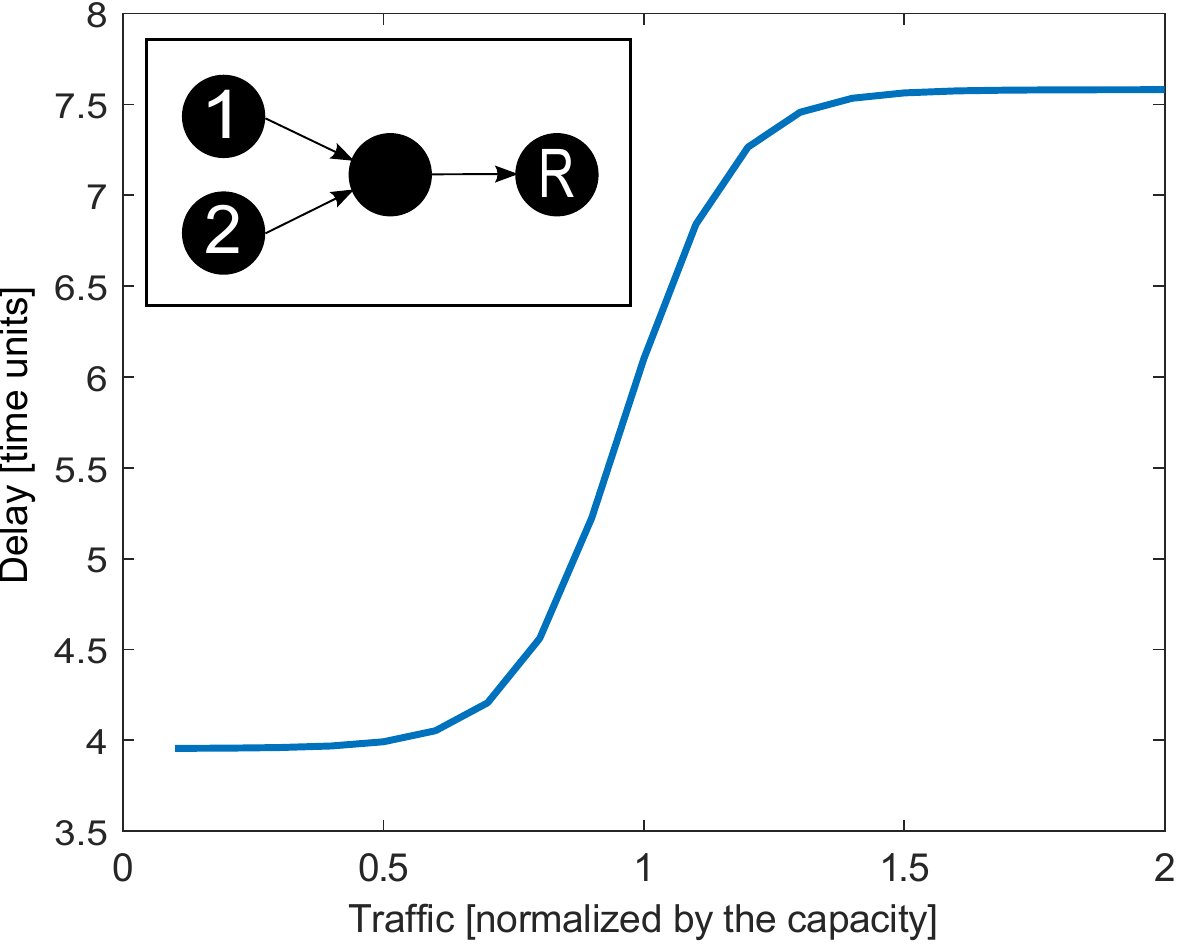}
  \caption{Simple example of the function to learn} 
  \label{fig:pStatement}
  \vspace{-0.25cm}
\end{figure}

Figure~\ref{fig:pStatement} shows the delay function we aim to learn (fit) with a NN in a very simple network topology (see inside the figure). In this example, only two nodes generate traffic which is sent to a third node. Specifically, the figure represents the average delay from $1$ to $R$ as a function of this traffic when the node $2$ is heavily massively traffic.

In real networks, this function becomes much more complex, since it depends on the state of the queues of the nodes of the path followed by each pair of nodes, which depends on the traffic sent among a big number of pair of nodes. In other words, Figure~\ref{fig:pStatement} exemplifies the function we aim to fit with NNs for a simple one dimension problem; however, in real networks, this is a high dimensionality function.

Specifically, the main questions that we aim to address in this paper are:

\begin{compactitem}
\item Can we train a NN to produce accurate estimates of the mean end-to-end delay for all pairs of nodes considering the input traffic matrix (ingress, egress)?
\item Which is the impact of fundamental network characteristics (topology, routing, size, traffic intensity) concerning  the accuracy of the NN?
\item Can we derive some guidelines to build the NN models? For example, what is the architecture of the NN model that best estimates of the delay?

\end{compactitem}

\section{State of the art}
\label{sec:soa}

The main goal of this paper is to understand if a NN can be used for network modeling %
and to provide practical guidelines for NN modeling in this scenario.

In the field of network modeling, there are two fundamental approaches: analytical and computational models (simulation) techniques. As for analytical techniques, Markov chain theory has been widely used in queuing theory to model the behavior of a single queue by assuming certain stochastic proprieties of the job arrival and job completion processes (ex. M/M/1, M/D/1...). These models have been extended to model networks of nodes, i.e., queuing networks~\cite{queuingTheory}. Examples of these theories are: Jackson Networks, Gordon-Newell theorem, Mean value analysis, Buzen's algorithm, Kelly network, G-network, BCMP network~\cite{queuingTheory2}. Computational models are also another popular technique to model the behavior of networks. Typically simulators (such as~\cite{inet}) operate either at packet or flow level and simplify the network protocols they simulate.

Machine Learning mechanisms have been used in the field of communications, and such techniques have been used extensively in the area of traffic analysis~\cite{trafficML}, network security~\cite{netSecurity} and root-cause analysis~\cite{RCA}. Additionally, some works propose the use of Reinforcement Learning techniques for routing optimization \cite{RLakyildiz}. In a recent paper~\cite{Wang}, the authors describe different use-cases for ML applied to network and among then, discuss NN as a modeling technique for computer networks.

Similarly to ~\cite{Wang}, in this paper, we advocate that NNs represent a third pillar in the field of network modeling. NNs can efficiently complement existing analytical and computational techniques providing important advantages. To the best of our knowledge, this is the \emph{first attempt to experimentally evaluate} the use of NNs to model the performance of computer networks.

\section{Methodology}
\label{sec:methodology}

In this section, we detail the methodology used to understand if a NN can estimate the average delay of a computer network as a function of the traffic matrix.

\subsection{Overview}

\begin{figure}
  \centering
  \includegraphics[width=\linewidth]{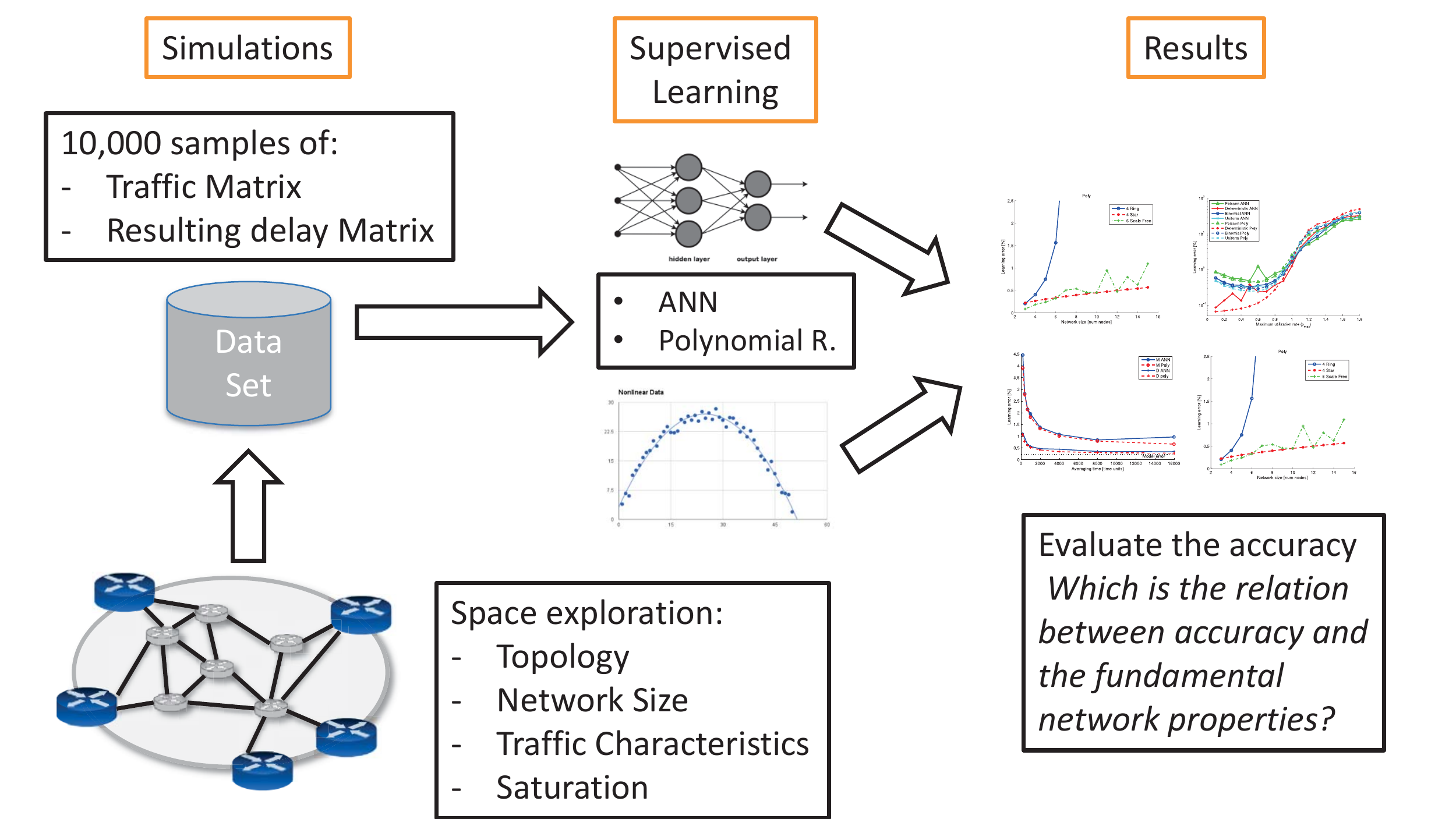}
  \caption{Scheme of the methodology followed in this work} 
  \label{fig:methodology}
  \vspace{-0.25cm}
\end{figure}

Figure~\ref{fig:methodology} shows an overview of the methodology. To experimentally analyze the accuracy of the NN, we generate different datasets by means of simulations, in each dataset we change different network characteristics: traffic distribution, traffic intensity, topology, size and routing policy and measure the average delay.

Once we have generated the dataset we use them to train a set of NN models, then we evaluate its accuracy using the cross-validation technique. We split the dataset into three sets, the training set with 60\% of the samples, the validation set with 20\% of the samples and the test set with the remaining 20\% of the samples. The training set is used to optimize the ML model, the validation set is used to evaluate the model during the training phase, and the test set is used to provide an independent evaluation of the performance. With this, we compare the average delay estimated by the NN model with the one measured from the simulator. To make a fair comparison, we subtract the unavoidable variance caused by the averaging process from each result.

Ultimately, we want to understand both the accuracy of the NN and its relation to fundamental characteristics of networks: traffic distribution, traffic intensity, topology, size and routing policy.

\subsection{Network Simulations}

In order to generate the dataset, we use the Omnet++ simulator (version 4.6)~\cite{omnet}, in each simulation we measure the average end-to-end delay during 16k units of time for all pairs of nodes. The transmission speed of all links in the network is set to 10 kilobits per unit of time, and the average size of the packets is 1 kilobits. 
We explore different network and traffic parameters to evaluate how these parameters affect the modeling capabilities when learning the network delay under different networks operating under different regimes of saturation and packet length.

Specifically, for the dataset we consider the following parameters:

\begin{compactitem}
\item \textbf{Topology:} We explore three different network topologies: unidirectional ring, star, and scale-free networks. These three topologies present different connectivities which may affect the learning capabilities.
\item \textbf{Network size:} We study networks from 3 to 15 nodes where all nodes are active transmitters and receivers.
\item \textbf{Traffic Distributions:} We evaluate four different packet length distributions: deterministic (constant), uniform, binomial and Poisson using a fixed average packet length. In all the cases the inter-arrival time is exponential.
\item \textbf{Traffic intensity:} We explore different levels of saturation in the network by varying the traffic intensity. For this, we transmit, among all pairs of nodes, a random value of traffic with a maximum value ($\rho_{max}$). We explore from very low saturated networks ($\rho_{max} = 0.1$) up to highly saturated networks ($\rho_{max} > 1$)
\item \textbf{Routing:} We explore three different routing configurations, which are detailed in section~\ref{sec:routing}.
\end{compactitem}

Overall we have generated more than 400 different datasets with different configurations in order to assess the accuracy of the NN. Each dataset consists of 10,000 different simulations (samples), and each sample contains the random traffic matrix, which is used as the input features in the ML model, and the delay matrix, which is used as the output features. Each dataset is divided into three sets to train, validate and test each model. All datasets used in this paper are publicly available~\footnote{https://github.com/knowledgedefinednetworking}.

\subsection{Neural Networks}

The generated dataset is used to train different neural networks (NNs) models, using the traffic matrices as input features and the delay matrices as output features. We explore the following NN hyper-parameters: number of hidden layers, number of neurons per layer, the activation function, the learning rate and the regularization parameter. We choose the hyper-parameters using the cross-validation technique and an independent test set to evaluate the accuracy of the model.

In terms of implementation, we use the Tensorflow library (version 1.2.1) to implement the NN models. After manual tunning of the hyper-parameters and unless noted otherwise, we use the following parameters: 

\begin{compactitem}
  \item Activation function: Sigmoid
  \item Number of hidden layers: Equal to the number of input, i.e., the square of the number of nodes in the network
  \item Maximum training epoch: 7,500,000
  \item Training Algorithm: Adam Optimizer
  \item Cost function: MSE with L2 regularization
  \item L2 regularization parameter: 0.00003
\end{compactitem}

In the results, we compute the accuracy of the models as ``learning error'' expressed as: 
\vspace{-0.1cm}
$$
  \label{eq:error}
  error =  \frac{1}{S}\frac{1}{N^2}\sum_{i=1}^{S}\sum_{i=1}^{N^2}(\hat{d}_i - d_i)^2 
$$
\vspace{-0.1cm}
where $\hat{d}_i$ is the predicted delay, $d_i$ is the test delay, $S$ is the size of the test-set and $N^2$ is the total number of pair of nodes in the network. To make a fair comparison, we subtract the unavoidable variance caused by the averaging process of each measure.

\section{Experimental Results}
\label{sec:experiments}

In this section, we present the results obtained in five different experiments that cover different synthetic network and traffic scenarios. Please note that unless stated otherwise, we only show exemplifying figures since the other cases provide similar results.

\subsection{Traffic Characteristics and Intensity}
\label{sec:saturation}

\begin{figure}
  \centering
  \includegraphics[width=0.9\linewidth]{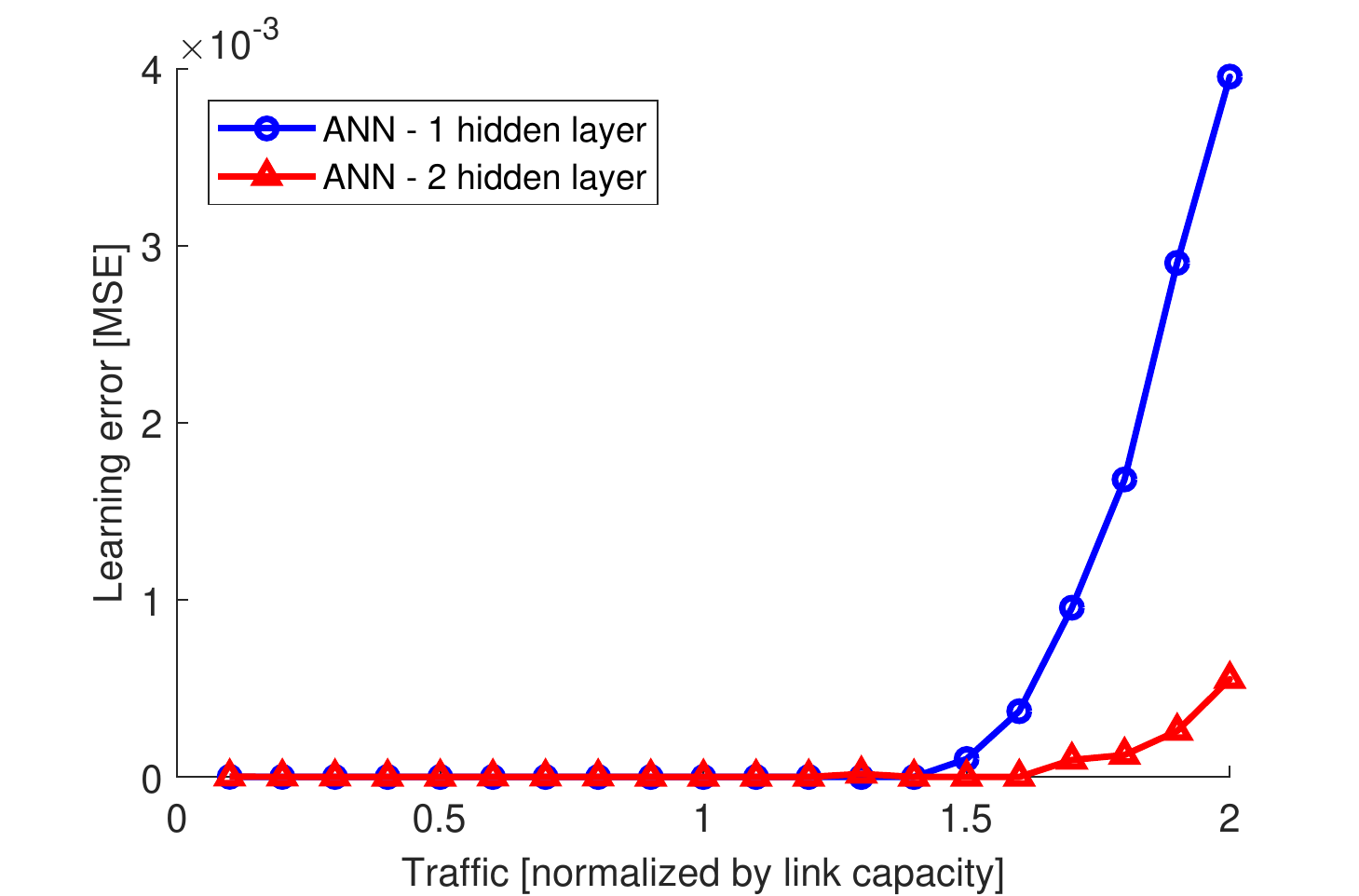}
  \caption{Learning error as a function of the network saturation (A total of 40 NN).} 
  \label{fig:saturation}
  \vspace{-0.25cm}
\end{figure}

First, we focus on the accuracy of the NN when estimating the delay of different traffic intensities and packet-size distributions. Figure~\ref{fig:saturation} shows the accuracy of two different NNs, with one and two hidden layers, in a 10-node scale-free network with different traffic intensity and for the binomial packet size distribution.

Please note that the traffic intensity is expressed as ($\rho_{max}$). Each pair of node generates a random uniformly distributed bandwidth with maximum ($\rho_{max}$). As an example ($\rho_{max}=2$) represents that, for each simulation, each node of the network, for each destination, generates a random traffic following the specified distribution at a random rate between 0 and the double of the link capacity divided by the number of destinations (i.e., uniformly distributed in the range of $(0, 2C]$, where $C$ is the link capacity). 

Figure \ref{fig:saturation} shows the learning error of two NN models with one and two hidden layers respectively. Both models exhibit a remarkably well performance, especially in low traffic scenarios. In high traffic scenarios, the function to learn becomes more complicated, and the deeper model clearly outperforms the smaller model. In the most saturated scenario, this error is roughly equivalent to a relative error of 0.7 \% for the deeper network. When the network is not saturated, in which the MSE is below $10^{-4}$, the relative error is practically negligible.

This experiment provides two interesting results. First, the fact that simple NNs do not perform well suggests that the delay function is complex and multi-dimensional, requiring sophisticated regression techniques such as deep NNs. In addition and more interestingly, deeper networks are required for saturated networks. This is because saturated networks result in more complex functions that require additional layers.

\subsection{Topologies and Network Size}

In this section, we explore the accuracy of the NN when estimating the delay with different network topologies and sizes.

\begin{figure}
  \centering
  \includegraphics[width=\linewidth]{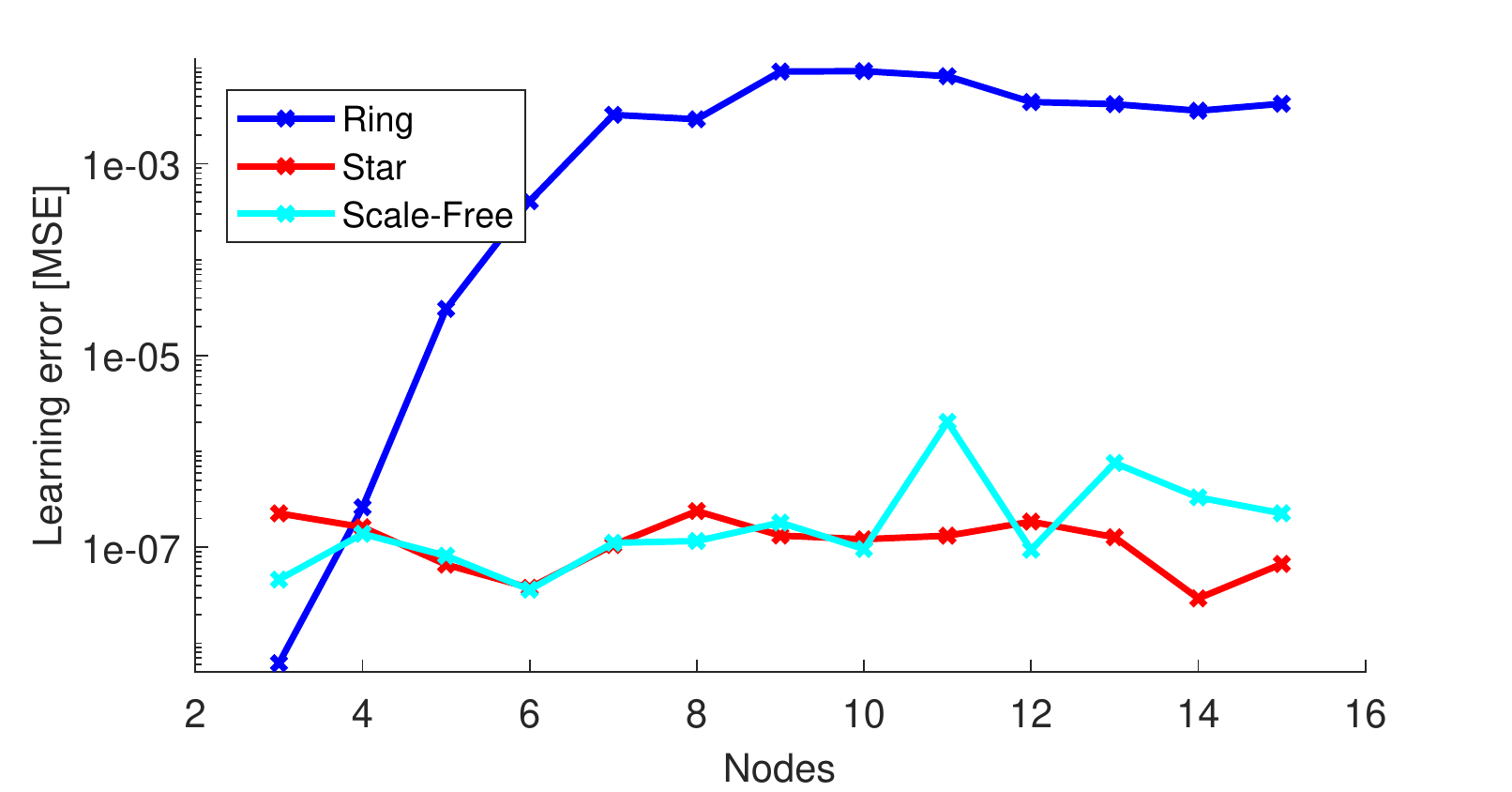}
  \caption{Learning error (log scale) as a function of the number of nodes for three different topologies (A total of 39 NN).} 
  \label{fig:topologies}
  \vspace{-0.25cm}
\end{figure}

Figure~\ref{fig:topologies} shows the accuracy of the NN model when estimating the delay in a ring, star and scale-free topology with different sizes, ranging from 3 to 15 nodes. In this scenario, the traffic intensity is set to ($\rho_{max} = 0.6$). As the figure shows, the NN model can accurately predict the delay in the star and the scale-free topology, but it presents a higher error in the ring scenario. 

The main reason for this is that for the same quantity of traffic, the topology determines the saturation of the network. In a ring topology, the traffic travels along the ring, which easily saturates the network. As a result, the delay is in the order of tens of seconds, and the shape of the function becomes more complex. In the scale-free topologies, and especially in the star topology, the intensity of traffic in the network is low and easy to model.

We conclude that, for the considered scenarios, the topology or the size \emph{has no impact} on the accuracy of the NN estimates. The only impact of such fundamental network characteristics is that depending on the specific topology this may increase the saturation, which results in more queuing requiring deeper networks as we have seen in the previous experiment.

\subsection{Routing}
\label{sec:routing}
In the third experiment, we explore the effect of the routing upon the learning capabilities. Table~\ref{tab:routing} shows the learning error when using three different routing policies in a scale-free network with 15 nodes and a traffic intensity of $\rho_{max} = 0.5$. For this, we consider the following routing policies: 'SP' correspond to the shortest-path approach, 'MAN' corresponds to manual designed routing policy that may not follow the shortest path approach but tries to load balance the use of the links. Finally, the 'POOR' configuration corresponds to a deliberately poor performance configuration, in which a few links are intentionally saturated.

In this case, we observe that, in the 'SP' and 'MAN' scenarios, the NN model performs well and as in the previous cases, they provide a low learning error. The reason is that the traffic intensity is low and easy to model. The 'POOR' routing policy performs worse since bottlenecks traffic through a few links that produce queuing saturating the network. 

In this set of experiments we have not seen \emph{any impact} of the routing policy in the accuracy of the NN. The only impact is again related to the queuing, if the routing policy produces saturation, then requires deeper models as we have shown in section~\ref{sec:saturation}.

\begin{table}[]
\def\arraystretch{1.35}
\centering
\caption{Learning error using different routings}
\label{tab:routing}
\vspace{-0.2cm}
\begin{tabular}{l|l|l|l|}
\cline{2-4}   & \multicolumn{1}{c|}{\textbf{SP}} & \multicolumn{1}{c|}{\textbf{MAN}} & \multicolumn{1}{c|}{\textbf{POOR}}  \\ \hline
\multicolumn{1}{|l|}{\textbf{Learning error}}  & $1.10\cdot 10^{-6}$  & $1.12\cdot 10^{-6}$  & $3.33\cdot 10^{-5}$   \\ \hline
\end{tabular}
\vspace{-0.35cm}
\end{table}

\subsection{Number of neurons per hidden layer}
\label{sec:neurons}

In the fourth set of experiments, we compare the size of the network, i.e., the nodes sending and receiving traffic with the number of neurons needed in the hidden layers of the NN models. Specifically, we explore three scale-free networks, with 5, 10 and 15 nodes and six numbers of neurons: 10, 25, 50, 100, 150, 225.

\begin{figure}
  \centering
  \includegraphics[width=\linewidth]{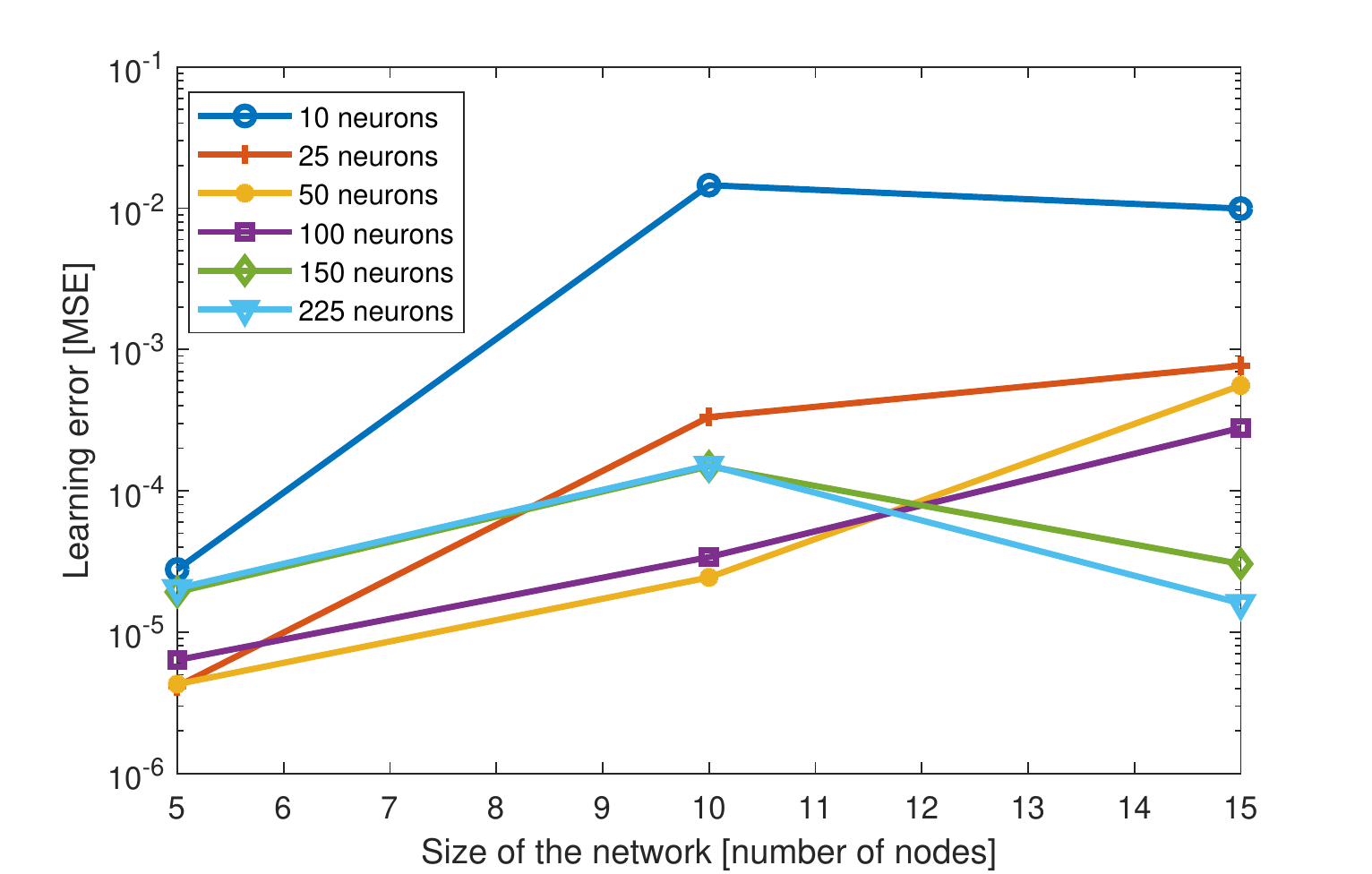}
  \caption{Learning error (log scale) for three different traffic intensities for a different number of neurons in the hidden layers (A total of 18 NN).} 
  \label{fig:neurons}
  \vspace{-0.25cm}
\end{figure}

Figure~\ref{fig:neurons} shows the learning error as a function of the number of nodes for a highly saturated network. We observe that, as the number of nodes increases, more neurons are needed to be able to model this behavior. This can be explained by the fact that the traffic in larger networks is multiplexed and demultiplexed several times, this increases the complexity and the dimensionality of the model. However, the use of more neurons in smaller networks is counter-productive since the model over-fits the training set, which increases the test error. To validate this claim, we have performed the same experiment in a low-traffic environment, and we have seen that in such environments fewer neurons are needed to train the NN adequately.

\subsection{Activation function}
\label{sec:neurons}
Finally and in the last set of experiments, we explore the effect of using different activation functions in the hidden layers. Specifically, we compare the sigmoid, hyperbolic tangent and rectified functions with the same set of hyper-parameters defined in section~\ref{sec:methodology}. In a low saturated network, the NN can learn the behavior of the network with a negligible error regardless of the activation function used. However in high traffic scenarios, the NN needs to model the behavior of the queues, and in such scenarios, we have obtained better results with the sigmoid activation function.

\section{Applying the guidelines to realistic environments}
\label{sec:realistic}

In order to apply the guidelines that we have learned with the synthetic experiments in this section, we train an NN in a realistic environment. Specifically, we use the GEANT2 24-node topology~\footnote{European optical transport network (www.geant.org)}. As for the traffic matrices we use a ``hot spot'' model~\cite{trafficMatrices}, where few pairs of nodes generate most of the traffic carried by the network.

\begin{table}[]
\def\arraystretch{1.35}
\centering
\caption{Learning error for three different traffic conditions in a realistic environment}
\label{tab:real}
\vspace{-0.2cm}
\begin{tabular}{l|l|l|l|}
\cline{2-4}   & \multicolumn{1}{c|}{\textbf{Low}} & \multicolumn{1}{c|}{\textbf{Medium}} & \multicolumn{1}{c|}{\textbf{High}}  \\ \hline
\multicolumn{1}{|l|}{\textbf{Learning error}}  & $1.60 \cdot 10^{-5}$  & $4.51\cdot 10^{-5}$  & $2.87\cdot 10^{-4}$   \\ \hline
\end{tabular}
\vspace{-0.35cm}
\end{table}

As for the NN in this set of experiments, we used two hidden layers, 576 neurons per layer (the square of the number of nodes), sigmoid activation function for the hidden layers, the MSE with L2 regularization as the cost function and the Adam optimizer algorithm.

Table~\ref{tab:real} shows the learning error in this scenario as a function of three different traffic intensities. We observe a similar behavior than in the synthetic experiments~\ref{sec:saturation}, and we obtain a good accuracy in all three cases with a relative error lower than 1\%.

\section{Discussion}
\label{sec:discussion}

In this section, we discuss our experimental results in order to have a better understanding of the use of neural networks (NNs) for computer network modeling. Given the high computational cost of the experiments depicted in this paper our results are limited to relatively small networks (up to 24 nodes), however, we conclude several valuable lessons:

\textbf{Neural networks can accurately model the average end-to-end delay as a function of the input traffic matrix for the considered scenarios:}
Well-designed NNS have produced excellent results in all cases. The main reason behind this is that, for the routing and forwarding mechanisms considered in the experiments, networks are deterministic systems with memory and can be learned with negligible error. As a result, NNs should be considered as a relevant tool in the field of network modeling.

\textbf{High traffic intensity requires deeper and bigger neural networks:} 
Modeling networks with low traffic can be modeled with simple tools. However, a valuable lesson is that networks that operate close to saturation require more sophisticated models, specifically more neurons and at least more than one layer. This is because the average end-to-end delay in a saturated network is a multi-dimensional non-linear function which requires deeper NNs. The saturation is the only parameter we have found that increases the complexity of the delay model. 
\newline

\subsection{Advantages and Disadvantages}

In this paper, we advocate that NNs are a useful tool in computer network modeling. In the following list, we discuss its main pros and cons with respect to more traditional modeling techniques:

\begin{compactitem}

\item \textbf{Accurate in complex scenarios:} Typically analytical models are based on strong simplifying assumptions of the underlying networking infrastructure: this is because they need to be tractable. On the other hand, simulations can model complex behavior, but this comes at a high development and computational cost. ML techniques and particularly NNs work very well with complexity (e.g., non-linear behaviors) and high-dimensionality scenarios.

\item \textbf{Fast and lightweight:} NNs require important computational resources for training, but once trained they are fast and lightweight. Indeed they can produce estimates of the performance of the network in one single step and require very little resources to run.
In other words, the training process implies iterating over all the dataset several times while back-propagating the error, whereas each evaluation of the model implies one forward-propagation step~\cite{ANNsurvey}.
This represents a significant advantage particularly in front of simulators that require important computational resources to run and might be slow.

\item \textbf{Data-driven models:} The main disadvantage of NNs is that they are data-driven techniques and as such require large training sets as well as computational resources for the learning phase. Additionally and in the context of computer networks collecting data from the network typically means sampling a distribution (for instance the average end-to-end delay needs to be sampled with several packets). This statistical measurement process is intrinsically associated with an irreducible error that impacts the learning accuracy. It is worth noting that in the experiments presented in this paper, we have subtracted this error to only focus on the error associated with the regressors.

\end{compactitem}

\section{Conclusions}
\label{sec:conclusions}

The main conclusion of this work is that the average end-to-end delay in communication networks can be accurately modeled using neural networks (NNs). We have found that NNs perform remarkably well, and the only handicap we have found is in highly saturated networks, which require deeper and more advanced NNs. NNs offer a powerful tool, and with a suitable tuning of the hyper-parameters, they can accurately model the average end-to-end delay of computer networks.

\section{Acknowledgments}
This work has been partially supported by the Spanish Ministry of Economy and Competitiveness under contract TEC2017-90034-C2-1-R (ALLIANCE project) that receives funding from FEDER and by the Catalan Institution for Research and Advanced Studies (ICREA).

\end{document}